\documentclass{iopart}
\usepackage{iopams}

\eqnobysec
\usepackage{graphicx}

\newcommand{\vecp}{\mathbf{p}}
\newcommand{\vecq}{\mathbf{q}}
\newcommand{\x}{\mathbf{x}}
\newcommand{\J}{\mathbf{J}}
\newcommand{\B}{\mathbf{B}}
\newcommand{\M}{\mathbf{M}}
\newcommand{\Id}{\mathbf{I}}
\newcommand{\opA}{\widehat{A}}
\newcommand{\opR}{\widehat{R}}
\newcommand{\opH}{\widehat{H}}
\newcommand{\opU}{\widehat{U}}
\newcommand{\opI}{\widehat{I}}
\newcommand{\doubleB}{\textbf{\textsf{B}}}
\newcommand{\doubleJ}{\textbf{\textsf{J}}}
\newcommand{\doubleN}{\textsf{N}}
\newcommand{\doubleId}{\textbf{\textsf{I}}}

\begin{document}
\title{Metaplectic sheets and caustic traversals in the Weyl representation}

\author{Alfredo M Ozorio de Almeida$^1$ and Gert-Ludwig Ingold$^2$}
\address{$^1$ Centro Brasileiro de Pesquisas Fisicas,
   Rua Xavier Sigaud 150, 22290-180, Rio de Janeiro, R.J., Brazil}
\address{$ ^2$ Institut f\"ur Physik, Universit\"at Augsburg,
   Universit\"atstra{\ss}e 1, D-86135 Augsburg, Germany}
\eads{\mailto{ozorio@cbpf.br} and \mailto{gert.ingold@physik.uni-augsburg.de}}

\begin{abstract}
The quantum Hamiltonian generates in time a family of evolution operators.
Continuity of this family holds within any choice of representation and, in
particular, for the Weyl propagator, even though its simplest semiclassical
approximation may develop caustic singularities. The phase jumps of the Weyl
propagator across caustics have not been previously determined.

The semiclassical appproximation relies on individual classical trajectories
together with their neighbouring tangent map.  Based on the latter, one defines
a continuous family of unitary tangent propagators, with an exact Weyl
representation that is close to the full semiclassical approximation in an
appropriate neighbourhood.  The phase increment of the semiclassical Weyl
propagator, as a caustic is crossed, is derived from the facts that the
corresponding family of tangent operators belong to the metaplectic group and
that the products of the tangent propagators are obtained from Gaussian
integrals.  The Weyl representation of the metaplectic group is here presented,
with the correct phases determined within an intrinsic ambiguity for the
overall sign. The elements that fully determine the phase increment across a
particular caustic are then analysed. 
\end{abstract}


\maketitle

\section{Introduction}

The construction of a path integral for the Weyl propagator
\cite{AMOA92,Report}, that is, the evolution operator, $\opU_t=\exp(-\rmi
t\opH/\hbar)$, in the Weyl representation, makes no restriction on the form of
the Hamiltonian that generates the continuous evolution of a quantum system.
Except for possible minor modifications due to ordering, the Weyl symbol,
$H(\x)$, for the quantum Hamiltonian, $\opH$, is identified with its classical
counterpart, $H(\bi{x})$, where $\{\bi{x}=(\vecp, \vecq)\}$ is the
$2N$-dimensional classical phase space, $\mathbf{R}^{2N}$. There is no need
for the classical Hamiltonian to be derived from a Lagrangian, so that the path
integral may just as easily be constructed for e.g. the Kerr Hamiltonian
\cite{WallsMil,TosValWis,Kirchmairetal}, $H(\bi{x})=(a{\vecp}^2 +
b{\vecq}^2)^2$, as for the usual form, $H(\bi{x})={\vecp}^2/ 2m +
V(\vecq)$. 

A classical trajectory coincides with a stationary phase of the Weyl path
integrand, just as with Feynman path integrals in the position representation.
Thus, one obtains semiclassical (SC) approximations for the full path integral
as a superposition of a few contributing classical trajectories.  The main
difference is that here the relevant trajectory is prescribed by a kind of
boundary condition on the {\it centre}, $\x \equiv (\bi{x}^+
+\bi{x}^-)/2$, of its pair of endpoints, $\bi{x}^-$ and $\bi{x}^+$,
rather than by the pair of end-positions $({\vecq}^-, {\vecq^+})$
\cite{Report,Ber89}. In both the Weyl and the position representations,
contributing trajectories may coalesce, with the passage of time, or with
continuous change in the boundary condition. At these {\it caustics}, the SC
approximation has a singularity.  In passing a caustic, SC approximations
generally change their phase by $\mu\pi/2$, where $\mu$ is known as the Maslov
index \cite{Maslov,MasFed}. If $H(\bi{x})={\vecp}^2/ 2m + V(\vecq)$, then,
in the position representation, $\mu$ coincides with the Morse index
\cite{Gutzwiller}.  A more general geometrical phase space treatment of the
Maslov index for periodic orbits is presented in references
\cite{Creaghetal90,Robbins91}.

The semiclassical approximation becomes exact in the limit where the
Hamiltonian is quadratic in the phase space variables. Classically, such a
Hamiltonian generates a (linear) {\it symplectic transformation} between the
endpoints of a classical trajectory: $\bi{x}^{-} \mapsto \bi{x}^{+} =
\M \bi{x}^{-}$, where $\M$ is a {\it symplectic matrix}. The symplectic
subgroup of classical canonical transformations, Sp($2N$), characterized by
$\M$, corresponds to the metaplectic subgroup, Mp($2N$), of general quantum
unitary transformations, $\opU_{\M}$, that is, the group of {\it metaplectic
operators}
\cite{Bargmann,KramMoshSel,GuilStern,Voros76,Voros77,Littlejohn86,deGosson06}.
The amplitude of the Weyl propagator for a metaplectic operator is constant
throughout phase space, but it has a true caustic singularity at an instant
when the SC form of the Weyl propagator must be replaced by a Dirac
delta-function.

The action of a Weyl propagator on other operators involves phase space
integrals.  Therefore, in the general SC scenario where the region of
integration may encompass different Maslov phases, it is crucial that these be
correctly evaluated.  Indeed, such switches of phase can be more important than
the possibly integrable SC singularities at the caustics themselves, as occurs
with the {\it initial value} or {\it final value} algorithms recently proposed
in \cite{OAValZam}. Furthermore, there are applications where SC {\it super
propagators} for Wigner functions \cite{Dittrich,AlmBro06} are specified by
products of Weyl propagators, each of which may traverse its own caustics
\cite{OAValZam,DitGoPa}, so that it is indispensible to determine the correct
Maslov phases in all cases.  The purpose of the present paper is to establish
the general connections of SC Weyl propagators (and of their Fourier
transforms) through their caustics.

In order to establish possible phase changes for caustic traversals, one needs
to focus on the {\it tangent map} for the full canonical transformation in the
neighbourhood of a given trajectory. In other words, one must study the
symplectic approximation to the full canonical transformation between the
neighbourhoods of the endpoints of a classical trajectory:
$\delta\bi{x}^{-} \mapsto \delta\bi{x}^{+} = \M_\x\;
\delta\bi{x}^{-}$, where the symplectic matrix for the transformations is
labelled by the centre, $\x = (\bi{x}^+ +\bi{x}^-)/2$, of the main
trajectory.  This is an essential ingredient of the SC approximation of the
Weyl propagator, $U(\x)$. Furthermore, one can also construct an exact
metaplectic {\it tangent operator}, $\opU_{\M_\x}$, represented by the Weyl
symbol $U_{\M_\x}(\x')$, i.e. the {\it tangent Weyl propagator}, based on the
same symplectic matrix, $\M_\x$. Just as the tangent map is close to the full
canonical transformation in the neighbourhood of its endpoints, we can define a
small neighbourhood $\delta \x'$, to be measured from the centre, $\x$, and
then we have $U_{\M_\x}(\delta\x') \approx U_{SC}(\delta\x')$.
\footnote[1]{One should recall that the Weyl representation is invariant with
respect to phase space translations, so that the change of origin in the
argument, $\x' \mapsto (\delta\x'=\x'-\x)$, of the SC approximation to $U(\x')$
is purely classical.} It is important to emphasize that this is a purely local
approximation and that different arguments of the Weyl propagator (i.e. choices
of the centre, $\x$) are related to different trajectories with different
linearized maps, so that the pure quantum tangent propagator is constructed
entirely from a local classical map.  Hence, one can investigate the possible
phase jumps of the SC approximation of the Weyl propagator by analyzing,
instead, the behaviour of the exact symbol for the tangent propagator.

The tangent Weyl propagators belong to the metaplectic group.  The difficulty
is that, even though Sp($2N$) and Mp($2N$) are isomorphic in the neighbourhood
of the identity, the correspondence $\M\mapsto\opU_\M$ is not one to one: There
is a double covering of the quantum group, i.e. a pair of metaplectic operators
and, hence, a pair of metaplectic Weyl propagators for each $\M$: ${\opU_\M}^-
= -{\opU_\M}^+$.  In particular, one should note that important particular
symplectic matrices, such as $\M=\Id$, the identity matrix, and $\M=-\Id$, the
reflection at the origin, correspond to pairs of metaplectic operators, even
though only one of these is commonly employed.  Since they only differ by an
overall sign, the choice of different sheets of a metaplectic operator has no
essential effect on how it acts on wave functions and this action completely
cancels out for Heisenberg operator evolution: $\opA \mapsto \opU^{\dag} \opA\;
\opU$. Nonetheless, the choice of sheet is essential for SC applications and it
lies in the focus of this paper, but we shall distinguish the sheet in the
notation for ${\opU_\M}$ only where it is essential.

The notion of {\it crossing a caustic} needs to be carefully dissected in terms
of corresponding families of classical maps and quantum operators: The quantum
Hamiltonian generates a continuous one-parameter family of evolution operators,
$\opU_t$, corresponding  semiclassically to a family of canonical
transformations, each of which may, in turn, be decomposed into its individual
classical trajectories.  In the neighbourhood of each of these trajectories,
one can then define a family of tangent symplectic maps, $\M_t$, and hence, a
continuous one-parameter family of tangent metaplectic operators:
${\opU}_{\M_t}$.  \footnote{Strictly, one should define $\M_{\x(t)}$ rather
than $\M_t$ following the previous notation, because the centre, $\x(t)$, of a
given trajectory with fixed initial condition is itself time dependent.
Indeed, one also deals with caustics as the centre $\x$ is varied while the
time is fixed, but one needs to consider only one-parameter families of
metaplectic operators, so this has been labelled as $t$ thoughout this paper.}
Such a family does not form a subgroup of the metaplectic operators, because it
is not closed with respect to multiplication, but the {\it product rule} for
each pair of sequential time intervals is satisfied: For $t_1 + t_2 = t$, one
has the exact product ${\opU}_{\M_{t_2}}{\opU}_{\M_{t_1}} = {\opU}_{\M_{t}}$,
corresponding to symplectic matrices, which also satisfy $\M_{t_2}\M_{t_1} =
\M_t$, even though neither $(\M_t)^2$ nor $(\opU_{\M_{t}})^2$ need belong to
their respective family.

Continuity of the classical family of symplectic transformations entails the
continuity of the family of tangent operators, but this crucial point will be
masked by caustics in any given representation. A caustic of the tangent Weyl
propagator occurs for the same matrix $\M_t$ as for the SC approximation of the
full evolution operator $\opU_t$, but it is a true singularity, at which the
Weyl symbol becomes a $\delta$-function \cite{Littlejohn86}.  The fact that
there is only a pair of metaplectic sheets, corresponding to the choice of a
$\pi$-phase (i.e. an overall sign) for the Weyl propagator, does not prevent a
possible phase change of $\pm\pi/2$ across a caustic. However, it will be shown
that it is only the overall sign that needs to be determined by the history of
each trajectory, because otherwise factors of $\rmi$ for tangent propagators
can be inferred from the dynamical properties of the symplectic matrix $\M_t$.

Fortunately, the caustics of different representations occur for different
values of the trace of $\M_t$ and thus at different parameter values $t$.  This
allows for swapping representations, where the continuity in a $t$-interval for
one representation supplies the sign change at the caustic of the other.  An
alternative method to evaluate the overall sign of the propagator beyond a
caustic is based on the exact product for sequential time intervals. These can
be chosen so that both propagators lie in the causticless neighbourhood, which
guarantees that the generalized Gaussian integral for their product, which lies
beyond the caustic, supplies the correct sign.

Nontrivial phase changes may in principle also occur for the product of finite
metaplectic operators, or for their product with translation (Heisenberg)
operators, or reflection (parity) operators.  The specification of these phases
is important for the SC treatment of operator and super operator evolutions,
such as \cite{OAValZam}. The case of parity operators demands special care
because they have been defined with different phases in different contexts,
namely the basis operator for the Weyl representation differs from  the
operators belonging to the metaplectic group that we consider here.

After a review of basic classical formulae for the Weyl propagators involving
symmetric matrices and generating functions in \sref{sec:classical}, the Weyl
representation of metaplectic operators and its Fourier transform, the chord
representation, are presented in \sref{sec:metaplecticops}. The choices which
are here made for the overall phases, that were ommited in \cite{Report}, is
justified by the general consistency of the propagators on switching
representation (\sref{sec:generalizedmaslov}) on the one hand, and from their
products (sections \ref{sec:producttransrefl} and
\ref{sec:productsmetaplectic}) on the other hand. 

Given the elegance with which the double sheeted topology of the metaplectic
group is here seen to be rendered within the Weyl representation, our aim is to
provide a pragmatic recipe for determining the phase increment across each
caustic, regardless of the metaplectic sheet. This is clearly synthetized in
the concluding \sref{sec:conclusions}. A didactical Appendix has been added to
illustrate the theory in the quintessential example of products of harmonic
oscilations.

\section{Classical ingredients}
\label{sec:classical}

The {\it centre generating function}, $S(\x)$, defines implicitly a canonical
transformation, $\bi{x}^- \mapsto \bi{x}^+$. This is achieved by a
finite version of Hamilton's equations, that is, defining the chord, $\bxi
\equiv \bi{x}^+ - \bi{x}^-$, we have 
\begin{equation}
\bxi = -\J \frac{\partial S}{\partial \x},\;\;\; \rm{or}\;\;\;\; 
\vecp^+ - \vecp^- = \bxi_\vecp = \frac{\partial S}{\partial \vecq}\;,
\;\;\; \vecq^+ - \vecq^- = \bxi_\vecq = -\frac{\partial S}{\partial \vecp},
\label{centran}
\end{equation}
where
\begin{equation}
\J = \left(
\begin{array}{cc}
     0 & -1 \\
     1 & 0 
\end{array}
\right) 
\end{equation}
in terms of $(\vecp,\vecq)$ blocks, that is, the standard symplectic matrix in
Hamilton's equations. The endpoints are given by
\begin{equation}
\bi{x}^{\pm} = \x \pm \frac{\bxi}{2}.
\end{equation}
For the canonical transformation generated by a Hamiltonian during a short
time, we thus have $S_t(\x) = -t H(\x)+ \mathcal{O}(t^3)$, where the third order
correction is given in \cite{Report}.

The tangent map for the canonical transformation near a trajectory centred on
$\x = (\bi{x}^+ + \bi{x}^-)/2$ is $\delta\bi{x}^{-} \mapsto
\delta\bi{x}^{+} = \M_\x\delta\bi{x}^{-}$.  It has the centre
generating function corresponding to the symplectic matrix $\M_\x$  
\begin{equation}
S_{\M_\x}(\delta\x) = \delta\x \cdot \B_\x \delta\x,
\label{centaction}
\end{equation}
where $\B_\x$ is the (symmetric) Hessian matrix for the full generating
function evaluated at $\x$. (Henceforth, the origin will be shifted so that
$\delta\x=\x$.)  Indeed, a symmetric matrix $\B$ defines the Cayley
parametrization of a symplectic matrix $\M$ \cite{Report}, and hence the
symplectic transformation $\bi{x}^+ = \M \bi{x}^-$, by
\begin{equation}
\M = (\Id + \J\mathbf{B})^{-1} (\Id - \J\mathbf{B}),
\label{Cayley1}
\end{equation}
with the inverse
\begin{equation}
\J\mathbf{B} = (\Id + \M)^{-1} (\Id - \M).
\label{Cayley-1}
\end{equation}
One should recall the fundamental symplectic property \cite{Arnold78},
\begin{equation}
\M' \J \M = \J,
\label{symplectic}
\end{equation}
where $\M'$ is the transpose of $\M$. 

It is important to note that any $2N$-dimensional symmetric matrix, $\B$,
defines a symplectic transformation through \eref{centran}, unless $\J\B$ has
an eigenvalue $-1$, whereas the condition  \eref{symplectic} for the symplectic
matrix $\M$ itself can only be verified a posteriori.  The eigenvalues of
$\J\B$ must come in pairs, $\pm \gamma$ just as the eigenvalues of
$\J\mathbf{H}$, which determine the symplectic flow for the Hamiltonian, $H(\x)
= \x \cdot \mathbf{H} \x/2$. They correspond to the same eigenvectors for which
$\M$ has a pair of eigenvalues $(\lambda, \lambda^{-1})$. Again, the fact that
the matrix $\B$ is real demands that any complex eigenvalues come in complex
conjugate pairs (see e.g. \cite{Arnold78}).  Notwithstanding that $\det(\B) =
\det(\J\B)$, one should keep in mind that the eigenvalues of $\B$ are
necessarily real, even if those of $\J\B$ need not be.

\begin{figure}
\centering
\includegraphics[width=0.45\textwidth]{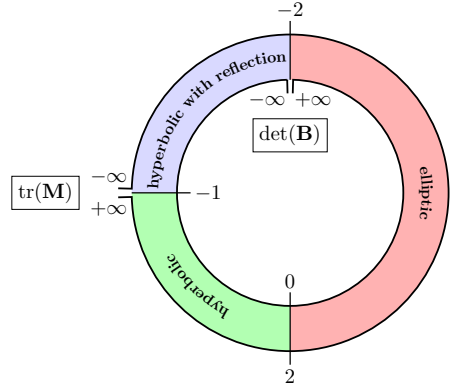}
\caption{Classification of elliptic and hyperbolic symplectic motion for a
single degree of freedom in terms of $\tr(\M)$ or $\det(\B)$.}
\label{Fig1}
\end{figure}

For $N=1$, the transformation can be classified by $\det(\B)$, that is, for
$\det\B >0$ the transformation is elliptic, for $-1<\det(\B)<0$ it is simply
hyperbolic and for $\det(\B)<-1$ hyperbolic with reflection.  One should note
that $\det(\B) = \det(\J\B) = \tr[(\J \B)^2]/2$.  The alternative
classification of symplectic transformations in terms of
\begin{equation}
\tr(\M) = 2\frac{1 - \det(\B)}{1 + \det(\B)},
\end{equation}
or of $\det(\B)$, is illustrated in \fref{Fig1}. The classification, for $N=2$
according to the invariants of $\J\B$ is given in \cite{RivSarOA} and, for
larger dimensional phase spaces, a similarity transformation can always reduce
the matrix $\J\B$, just as the symplectic matrix $\M$, into 2-dimensional or
4-dimensional blocks \cite{Arnold78,Williamson}.

A continuous transition between elliptic and hyperbolic transformations may
pass through the identity, but, in general the boundary between them lies along
the phase space shears, for which the normal form will be Cayley-parametrized
by   
\begin{equation}
\label{Bshear}
\B \rightarrow 
\left(
\begin{array}{cc}
     b & 0 \\
     0 & 0 
\end{array}
\right) \;, \;
\J\B \rightarrow 
\left(
\begin{array}{cc}
     0 & 0 \\
     b & 0 
\end{array}
\right) \;,
\end{equation}
for some real parameter $b$, such that $b=0$ for the identity.  This is a
parabolic transformation, such that its symplectic matrix has the form of a
Jordan block:
\begin{equation}
\label{Mshear}
\M \rightarrow \left(
\begin{array}{cc}
     1   & 0 \\
     -2b & 1 
\end{array}
\right) \;.
\end{equation}

The breakdown of the Cayley parametrization \eref{Cayley1}, such that $\det(\B)
\rightarrow \pm\infty$, for a one-parameter family of transformations, if
$N=1$, is specified by the limiting matrix,
\begin{equation}
\label{Mrefl}
\M \rightarrow \left(
\begin{array}{cc}
     -1 & 2b' \\
      0 & -1 
\end{array}
\right) \;,
\end{equation}
so that the family of symplectic transformations goes through a shear with a
reflection through the origin.  This is the generic form for a caustic, i.e.
for a single degree of freedom we have $\tr(\M) = -2$, or equivalently,
$\det(\Id + \M) = 0$. But there is only one eigendirection, unless the
parameter $b'=0$ in \eref{Mrefl}, in which case the transformation reduces to a
reflection at the origin.

The neighbourhood of a caustic can be neatly treated with the aid of the
complementary Cayley parametrization \cite{Report,deGosson06},
\begin{equation}
\M = -(\Id - \J\tilde{\mathbf{B}})^{-1} (\Id + \J\tilde{\mathbf{B}}),
\label{Cayley2}
\end{equation}
with its inverse:
\begin{equation}
\J\tilde{\B} = - (\Id - \M)^{-1} (\Id + \M).
\label{Cayley-2}
\end{equation}
From \eref{Cayley1} and \eref{Cayley2} one obtains the explicit expression for
the new Cayley matrix as
\begin{equation}
\tilde{\B} = -\J{\B}^{-1}\J = {\J}^{-1}{\B}^{-1}\J.
\label{BB} 
\end{equation}
Thus, $\B^{-1}$ and $\tilde{\B}$ are related by a similarity transformation and
can be diagonalized simultaneously, so that $\B$ and $\tilde{\B}$ have the same
signature and their determinants have the same sign. It follows that
$\det(\tilde{\B})\rightarrow 0$ at a caustic, which allows for the
neighbourhood of the caustic to be described by
\begin{equation}
\label{Bshearefl}
\tilde{\B} \rightarrow 
\left(
\begin{array}{cc}
     \tilde{b} & 0 \\
     0         & 0 
\end{array}
\right) \;, \;
\J\tilde{\B} \rightarrow 
\left(
\begin{array}{cc}
     0         & 0 \\
     \tilde{b} & 0 
\end{array}
\right) \;,
\end{equation}
such that $\tilde{b}=0$ for a pure reflection. Thus, $\det(\B)$ is small near
the identity and diverges near the reflection, whereas the situation for
$\det(\tilde{\B})$ is the reverse. Just as the Cayley matrix $\B$ specifies a
quadratic centre generating function,
\begin{equation}
{\tilde {S}}_\M(\bxi) = \frac{1}{4}\bxi\cdot\tilde{\B}\bxi
\label{chordgen}
\end{equation}
is the {\it chord generating function} for a symplectic transformation
\cite{Report}, such that $\x = \J\partial \tilde{S}/ \partial \bxi$ and again
$\bi{x}^\pm = \x \pm \bxi/2$. Thus, whereas $S(\x)$ becomes singular at a
reflection, because all the chords then have the same centre, $\tilde{S}(\bxi)$
becomes singular for a uniform translation, for which all centres have the same
chord.

One should note that, since the eigenvalues of both $\J\B$ and $\J\tilde{\B}$
come in pairs of opposite sign, it is generally true that $\tr(\J\B) =
\tr(\J\tilde{\B}) = 0$.  Furthermore, one obtains from \eref{Cayley-1}
and \eref{Cayley-2} that the eigenvalues for $\J\tilde{\B}$ and $\J\B$ are
related by $\tilde{\lambda}=-\lambda^{-1}$.  Hence, $\det(\B)$ and
$\det(\tilde{\B})$ change signs simultaneously at the boundary between hyperbolic
and elliptic transformations, so that the classification in \fref{Fig1} also
holds for $\tilde\B$, providing that one interchanges $0\leftrightarrow\infty$.
If $N=1$, both signatures $\sigma(\B)=\pm2$ and $\sigma(\tilde{\B})=\pm2$ for
an elliptic transformation, whereas $\sigma=0$ for both parametrizations in the
case of a hyperbolic transformation.

\section{Metaplectic operators in the Weyl and the chord representations}
\label{sec:metaplecticops}

For the present purpose, the most appropriate way to consider the Weyl
representation of an operator $\hat{A}$, i.e. its \textit{centre symbol} or
\textit{Weyl symbol}, is
\begin{equation}
\label{covW} 
A(\x) = 2^N\tr\left(\hat{R}_{\x}\hat{A}\right),
\end{equation}
where $\hat{R}_{\x}$ is the quantum operator corresponding to a reflection
through the phase space point $\x$ \cite{Report,Grossmann,Royer}. Within a
translation, this is just the symplectic transformation with $\M=-\Id$,
$\tilde\B=\mathbf{0}$, though its main Cayley matrix $\B$ has a singular
limit.  It is important to note that such an operator is only defined within an
overall phase, but in all standard definitions, one imposes that
$(\hat{R}_{\x})^2=\opI$. 

The metaplectic operator, corresponding to the symplectic transformation
$\bi{x}^+ = \M \bi{x}^-$ and Cayley matrix $\B$, has as Weyl symbol
the {\it Weyl propagator}, 
\begin{equation}
\fl U_\M(\x) 
\equiv \pm \frac{2^N}{[\det(\Id+ \M)]^{1/2}}
\exp\!\left(\frac{\rmi}{\hbar}\x \cdot \B \x\right)
= \pm [\det(\Id + \J \B)]^{1/2}
\exp\!\left(\frac{\rmi}{\hbar}\x \cdot \B \x\right),
\label{Uweyl}
\end{equation}
which was derived in \cite{Report} only within a phase. Thus, it should be
noted that both amplitudes have here been specified by the square root of a
determinant, rather than that of its modulus.  There will then be a factor of
$\rmi$, if the determinant is negative. In other words, the duplicity of the
metaplectic sheet coincides with that of the Riemann sheet for the square root,
as shall be verified further on.  The $(\pm)$ signs which distinguish the pair
of sheets are kept as a reminder that either choice can result for a
metaplectic operator labelled by the same $\M$, depending on its previous
evolution. \footnote{Other representations of the metaplectic operators, such
as the usual position propagator, are also exact in their SC form. One can
identify $\M \mapsto \opU_\M$ in all cases, but the special appeal of the
present phase space representations is that they are specified by invariants
of the full corresponding symplectic matrix, $\M$.}
 
One should keep in mind that for general motions the only difference between
this tangent propagator and the approximate SC Weyl propagator, corresponding
to a nonlinear transformation, is that the action of the former is obtained as
the homogeneous second order approximation to the full action $S(\x)$. (It
follows that the amplitude of the tangent propagator is constant.) In other
words, \eref{centaction} may either be interpreted as the local phase of the SC
propagator, or the exact phase of the tangent propagator.  Thus, in the
neighbourhood of the origin, which corresponds to the midpoint of a trajectory
for the full nonlinear motion, one can identify the overall sign of both
propagators.

The alternative forms for the determinantal amplitudes are immediately
interpreted with the aid of the transformations between midpoints and
endpoints:
\begin{equation}
\x = \frac{1}{2} (\Id+\M)\bi{x}^- \;\; \rm{or} \;\; \bi{x}^- = (\Id+ \J\B)\x,
\label{initialmid}
\end{equation}
whereas
\begin{equation}
\x = \frac{1}{2} (\Id+\M^{-1})\bi{x}^+ \;\; \rm{or} \;\; \bi{x}^+ = (\Id- \J\B)\x.
\end{equation}
Thus, the Jacobians for these transformations are
\begin{equation}
\det(\Id\pm \J\B) = 2^N\det(\Id+\M)^{-1} = 2^N\det(\Id+\M^{-1})^{-1},
\label{Jacobis} 
\end{equation}
which also allows for the expression of the amplitude in terms of $\det(\Id-
\J\B)$, as in \cite{Report}.  In the present text, the choice of sign within
the determinant in the LHS of \eref{Jacobis} is never altered in order to avoid
any confusion with the issue of determining the overall Maslov phases. 

Any continuous family of symplectic transformations corresponds to a continuous
family of metaplectic operators. If these are represented by their Weyl symbol,
then the latter must also be continuous, except at the caustics, where
$\det(\B)\rightarrow \infty$.  Recalling that the Weyl representation of the
operator $\opI$ is just the constant, $I(\x)=1$, it follows that all the
metaplectic operators that are continuously connected to $\opI$ are correctly
described by the $(+)$ sign in \eref{Uweyl}.  It will be verified further along
that the pair of metaplectic sheets is correctly accounted for, including
caustic traversals, by the square roots of these determinants, without taking
their modulus. 

The Weyl propagator for the harmonic oscillator $(N=1)$, driven by the
Hamiltonian $H(x) = (\omega/2) (p^2 + q^2)$, provides an iluminating example.
After a time $t$, the Cayley matrix for this typical elliptic propagator will
be $\B(t) = -\tan(\omega t/2) \Id$, so that the full Weyl propagator is
\begin{equation}
U_t(\x) = \frac{1}{\cos(\omega t/2)} 
\exp\!\left[-\frac{\rmi}{\hbar} \tan (\omega t /2)(p^2 + q^2) \right].
\label{Uho}
\end{equation}
One should note the simple form taken by the square root in the amplitude of
\eref{Uweyl}. The Maslov theory in the next section will confirm that the
change of phase across the caustic is just that shown in \eref{Uho}, in which
the sign ambiguity is fully accounted for by adding an even or odd factor of
$2\pi$ to the cosine phase. So there is a minus sign as the boundary $|\omega
t|=\pi$ is crossed, coinciding with the sign of the cosine. This signifies a
change of metaplectic sheet. At $\omega t=2\pi$, one obtains $-\opI$, that is,
$U(\x)=-1$ and it is only after a second rotation in phase space that the
identity operator is regained. In the absence of the history of the evolution,
the $(\pm)$ sign is undetermined, but our general concern are the particular
phase jumps for particular caustic transitions. Products of metaplectic
operators for the harmonic oscillator are discussed in the Appendix.

In contrast, for the inverted harmonic oscillator, $H(x) = (\lambda/2) (p^2 -
q^2)$, the Weyl propagator is
\begin{equation}
U_t(\x) = \frac{1}{\cosh(\lambda t/2)} \exp\!\left[-\frac{\rmi}{\hbar}
\tanh(\lambda t /2)(p^2 - q^2) \right].
\label{Uiho}
\end{equation}
In the case of such a quadratic hyperbolic Hamiltonian, there is no caustic
singularity. Above, we have portrayed the propagators continuously connected to
$\opI$, whereas those in the neighbourhood of $-\opI$ would all have a
negative sign. The hyperbolic transformations with reflection (see
\fref{Fig1}) cannot be reached continuously through the action of a
hyperbolic Hamiltonian.  Even so, to complete the types of symplectic
transformation portrayed in \fref{Fig1}, one may add the continuous family
of operators (for $t$ positive or negative):
\begin{equation}
U_t(\x) = \pm\frac{\rmi}{\sinh(\lambda t/2)} \exp\!\left[-\frac{\rmi}{\hbar}
\coth(\lambda t /2)(p^2 - q^2) \right].
\label{Uriho}
\end{equation}
However, it should be noted that the corresponding hyperbolic transformations
with reflection are continuously obtained from the reflection, for which
$\M=-\Id$ and $\tilde\B=0$, while they cannot be reached directly from $\Id$,
without traversing a caustic.  For this reason the phase presented in
\eref{Uriho} anticipates the theory in the following sections.  One should note
that typical metaplectic operators for elliptic and hyperbolic transformations
are simply obtained from these examples, owing to the invariance of the Weyl
representation with respect to symplectic transformations. 

The Cayley matrix for the harmonic oscillator becomes singular for $\omega t =
\pi$, when the transformation is just a reflection.  The correct corresponding
Weyl propagator is a Dirac delta-function, which is indeed singular.  In
general, a continuous family of metaplectic operators has a singular Weyl
representation as $\det(\Id+\M) \rightarrow 0$. If $N=1$, this is equivalent to
$\tr(\M) + 2 \rightarrow 0$.

The chord representation of an operator $\hat{A}$, i.e. its {\it chord symbol},
is
\begin{equation}
\label{covC} 
\tilde{A}(\bxi) = \tr\left(\hat{T}_{-\bxi}\;\hat{A}\right) ,
\end{equation}
where $\hat{T}_{\bxi}$ is a Heisenberg operator, that is, the unitary operator
corresponding to a phase space translation by the vector $\bxi$. (It is referred
to in various ways throughout the literature, but here we follow the notation
in \cite{Report}.) This is the conjugate representation to the Weyl
representation, so that the {\it chord propagator} corresponding to the same
metaplectic operator and, hence, the same classical symplectic matrix, $\M$,
can be evaluated through the symplectic Fourier transforms (see e.g.
\cite{Report}),
\begin{equation}
 \tilde{U}(\bxi) = \frac{1}{(2\pi\hbar)^N}\int\rmd\x 
 \exp\!\left(\frac{\rmi}{\hbar}\x\wedge \bxi\right)
U(\x)
 \ ,
\label{FWS}
\end{equation}
\begin{equation}
U(\x) = \frac{1}{(2\pi\hbar)^N}\int\rmd\bxi
\exp\!\left(\frac{\rmi}{\hbar}\bxi\wedge \x\right)
\tilde{U}(\bxi),
\label{FCS}
\end{equation}
where we recall that the {\it wedge product} $\x\wedge\bxi \equiv
\bxi\cdot\J\x$.

To obtain the metaplectic chord propagator, one takes the Fourier transform of
the complex Gaussian form of the Weyl propagator \eref{Uweyl}, 
\begin{equation}
{\tilde U}_\M(\bxi) = \pm[\det(\Id + \J\B)]^{1/2}
\int \frac{\rmd\x}{(2\pi\hbar)^{N}}
\exp\!\left[\frac{\rmi}{\hbar} (\x \cdot\B\x - \x \cdot\J\bxi)\right]
\label{tanchord}
\end{equation}
and uses \eref{BB} so as to express the chord propagator in terms of the
alternative Cayley matrix, $\tilde\B$, defined by \eref{Cayley-2}. Then the
amplitude of the chord propagator results from the equality
\begin{equation}
\det(\Id+ \J{\tilde\B}) = \det(\Id+\J\B)\det(\B)^{-1}.
\label{chordamp}
\end{equation}
If the number of negative eigenvalues of the $2N$-dimensional matrix $\B$ is
$N_-$, then its signature is $\sigma(\B)= 2(N-N_-)$, so that the Gaussian
integral gains the phase $\pi (N-N_-)/2$.  On the other hand, one can also
express $\arg[\det(\B)]= \pi [N_-(\mathrm{mod}\; 2)]$, so that the new phase can be
incorporated into $\det(\Id+ \J{\tilde\B})^{1/2}$, within a phase $\nu\pi$,
for some integer $\nu$. This leads to \footnote{Note that this corrects the
sign of the exponent in (6.40) of reference \cite{Report}.}
\begin{eqnarray}
\label{Uchord}
 {\tilde{U}}_\M(\bxi) 
 &= \pm(\rmi)^N [\det(\Id+\J\tilde\B)]^{1/2}
    \exp\!\left(-\frac{\rmi}{4\hbar}\bxi\cdot \tilde\B\bxi\right) \\ \nonumber
 &= \pm \frac{(2\rmi)^N}{[\det(\Id-\M)]^{1/2}}
    \exp\!\left(-\frac{\rmi}{4\hbar}\bxi\cdot \tilde\B\bxi\right), 
\end{eqnarray}
where the overall $(\pm)$ sign need not coincide with the original Weyl
representation (i.e. positive near the identity) and all further phases are
specified by the square roots of the determinant.  One can also interpret both
forms of the amplitude in terms of the relation between the endpoint and the
chord:
\begin{equation}
\bxi = [\M-\Id]\bi{x}^- \;\; \rm{or} \;\; \bi{x}^- =\frac{1}{2} [\J\tilde\B-\Id]\bxi.
\label{initialchord}
\end{equation}

In practice, the increment in phase resulting from the Fourier transform, 
\begin{equation}
\Theta=\frac{\pi}{4} \sigma(\B)= \frac{\pi}{2}(N-N_-) ,
\end{equation}
is just added onto the overall preexisting phase in the Weyl representation
(i.e. zero, in a causticless neighbourhood of the identity), so as to determine
the overall phase in \eref{Uchord}.  For example, one may perform directly the
symplectic Fourier integral on \eref{Uho} for the harmonic oscillator.  The
overall sign for the Weyl propagator is (+) for $\omega t<\pi$, so the choice
of sign for the chord propagator will correspond to the phase $(\pi/4)
\sigma(\B) =-(\pi/2)\omega t/ |\omega t|$, which leads to
\begin{equation}
\tilde{U}_t(\bxi) = -\frac{\rmi}{2\sin(\omega t/2)}\exp\!\left[\frac{\rmi}{4\hbar}
                     \cot(\omega t /2)({\bxi_p}^2 + {\bxi_q}^2) \right].
\label{Uhochord}
\end{equation}
Likewise, the Fourier transform for the inverted oscillator \eref{Uiho} leads
to the chord propagator:
\begin{equation}
\tilde{U}_t(\bxi) = \frac{1}{2\sinh(\lambda t/2)}\exp\!\left[\frac{\rmi}{4\hbar}
                    \coth(\lambda t /2)({\bxi_p}^2 + {\bxi_q}^2) \right].
\label{Uihochord}
\end{equation}
whereas the reflected hyperbolic chord propagator, obtained from \eref{Uriho},
becomes
\begin{equation}
\tilde{U}_t(\bxi) = -\frac{\rmi}{2\cosh(\lambda t/2)}\exp\!\left[\frac{\rmi}{4\hbar}
                    \tanh(\lambda t /2)({\bxi_p}^2 + {\bxi_q}^2) \right].
\label{Urihochord}
\end{equation}

Here, one should remark that both the oscillator, for $\omega t\rightarrow\pi$,
and the reflected hyperbolic family for $\lambda\rightarrow 0$ correspond to a
reflection about the origin, but this operator does not have the same phase as
the operator $\hat{R}_{\x}$ employed in the definition of the Weyl
representation. Indeed, the chord symbol of the standard reflection is
${R}_{\x}(\bxi)=2^{-N}$, so that one should distinguish the present {\it
metaplectic reflection} as $\hat{R}'_{\x}= \rmi^{-N}\hat{R}_{\x}$ (i.e.
$R'_\x(\bxi)=(2\rmi)^{-N}$), which satisfies  $(\hat{R}'_{\x})^2=(-1)^N \opI$.
This subtle phase distinction has surely been anticipated in other contexts
(e.g. in \cite{Boiuna10}), but it is for manipulations in the Weyl
representation that it is of crucial importance.

Returning to the full expression for the metaplectic chord propagator
\eref{Uchord}, one realizes that the limit $\tilde\B\rightarrow 0$ must specify
the metaplectic reflection, $\pm\hat{R}'_{0}$. Indeed, all operators that are
continuously connected to such a reflection will preserve the same overall sign
and the boundary of this region is only reached as one of the eigenvalues of
$\tilde\B$ reaches the value $+1$.  Thus there is no caustic for the chord
representation as an eigenvalue of $\tilde\B$ goes through zero, even though
there may be a switch of type (elliptic$\leftrightarrow$hyperbolic, if $N=1$).
In the previous examples, one thus verifies that the transition between the
chord propagator for the harmonic oscillator (as $\omega t \rightarrow \pi$)
and the hyperbolic with reflection (as $\lambda t \rightarrow 0$) is smooth.
Furthermore, all chord propagators within a causticless neighbourhood of the
reflection, $\hat{R}'_{0}$, corresponding to the phase $\pi$ of the harmonic
oscillator, share the overall phase $-\pi/2$.

This scenario is just the complement of that previously observed for the Weyl
representation, in which all operators that are continuously connected to the
identity must share the same overall sign.  In other words, the continuous
group of metaplectic operators is represented with a cut in the Weyl
representation that includes the reflection, while its cut in the chord
representation includes the identity. Each representation is smooth along the
other's singularity.  This is the basis of the Maslov method of phase
determination in the following section.
 
The alternative approach to determine the Maslov phases is to rely on products
of metaplectic operators.  At the passage of a one-parameter family of tangent
operators through the singularity of its Weyl representation, an operator,
characterized by large $|\det(\B)|$ or small $|\det(\tilde\B)|$, is multiplied by
an operator close to the identity, i.e. with small $|\det(\B)|$. Thus, the
determination of the final overall sign should be included in the problem of
establishing the overall phase for the product of any pair of metaplectic
operators. This is obtained from the product rule in the Weyl representation
for arbitrary pairs of operators \cite{Report}, $\opA = \opA_2 \opA _1$:
\begin{equation}
\fl [A_2 A_1](\x) = \int \frac{\rmd\x_2\rmd\x_1}{(\pi \hbar)^{2N}} A_2(\x_2) A_1(\x_1)
\exp\!\left[\frac{2\rmi}{\hbar} (\x_2 - \x) \cdot \J(\x_1 - \x)  \right].
\label{prule}
\end{equation}
In the case of metaplectic operators, each Weyl propagator is a complex
Gaussian, so that \eref{prule} is a $4N$-dimensional Gaussian integral. This
is reduced to a $2N$-dimensional integral in the limiting case where either
$\B=0$ or $\tilde\B=0$, for one of the factors (i.e. it is either a translation
or a reflection). These special cases are considered in
\sref{sec:producttransrefl}.

\section{Generalized Maslov method for the overall sign of the propagator}
\label{sec:generalizedmaslov}

Continuity of the family of tangent operators (corresponding to the
neighbourhood of a trajectory with a given initial value) implies the
continuity of the family of tangent Weyl propagators in a causticless
neighbourhood of the identity operator.  Hence, the overall sign of the SC
approximation to the full Weyl propagator must be preserved in this region,
regardless of the type of symplectic transformation (hyperbolic or elliptic, if
$N=1$): A transition between these two types within this neighbourhood, at a
tangent propagator characterized by \eref{Bshear}, does not alter the overall
sign.

This scenario is in marked contrast to the chord propagator in this same
region, which has its caustic exactly for those transformations characterized
by \eref{Bshear}.  The change of phase across this boundary is then evaluated
by just the procedure followed in the previous section, that is, by performing
the Fourier transform from the Weyl propagator.  This is a generalization of
the Maslov method \cite{Maslov} that has already been tacitly employed to
obtain the change of sign for the harmonic oscillator \eref{Uhochord} at $t=0$.
\footnote{In the double phase space scenario for semiclassical approximations,
the evolution operator corresponds to a Lagrangian surface expressed in terms
of the conjugate centre or chord coordinates \cite{AlmBro06}. Thus, one obtains
complete equivalence to switching between positions and momenta in ordinary
phase space.} 

The same method can now be employed to explore the possible change of phase
within the Weyl representation across its caustic, characterized by
\eref{Bshearefl}.  What then is the result of reversing the Fourier
transformation, so as to return to the tangent Weyl propagator from the chord
propagator?  There is certainly no change if no caustic has been crossed, since
this procedure may be continuously connected to the simple task of performing
and then reversing the same Fourier transformation.  In terms of Gaussian
integrals, this general conclusion follows from the important result in
\sref{sec:classical}, that both Cayley matrices for a given transformation
$\M$, i.e. $\B$ and $\tilde\B$, have the same signature.

Let us now consider a one-parameter family of tangent operators, which evolves
from an original operator, characterized by $(\M, \B, \tilde\B)$, such that
$\sigma(\B)=\sigma(\tilde\B) = 2(N-N_-)$, to a new operator, characterized by
$(\M', \B', \tilde\B')$, with $\sigma(\B')=\sigma(\tilde\B') = 2(N-N'_-)$.  If
a single Weyl caustic is crossed, the continuity of this operator evolution is
still maintained within the chord representation, so that there is no change of
the overall phase of the chord propagator.  The increment of the overall phase
of the Weyl propagator will then just depend on the pair of Fourier integrals:
\footnote{Notice that the exponents of the Weyl propagator and the chord
propagator have different signs.}
\begin{equation}
\Theta = \frac{\pi}{4} [\sigma(\B')-\sigma(\B)] = \frac{\pi}{2}(N_- -N'_-). 
\label{causticphase}
\end{equation}
There is no phase increment if $N'_-=N_-$. Alternatively, $\Theta$ need not
equal $\nu\pi$ (with integer $\nu$), if the evolution has changed the type of
the transformation, being that the square roots in the amplitudes may be
imaginary. This is just the case, if $\M$ denoted an elliptic transformation
(harmonic oscillator \eref{Uho}), whereas $\M'$ becomes a reflected hyperbolic
transformation \eref{Uriho}, though both transformations are smoothly connected
in the chord representation.

In general, the sign of the determinant in the final Weyl propagator
\eref{Uweyl} already picks up the signatures for the pair of Fourier
transformations, so as to specify the phase of the Weyl propagator within $\pm
\pi$. That is, using \eref{chordamp} and its reverse, as well as continuity of
the chord propagator, one obtains 
\begin{eqnarray}
\arg[\det(\Id+ \J{\B'})] &= \arg[\det(\Id+ \J\tilde\B')]-\arg[\det(\tilde\B')] \\ \nonumber
&=\arg[\det(\Id+ \J\B)]-\arg[\det(\B)]+\arg[\det(\B')],
\end{eqnarray}
so that the  phase increment due to the determinant is just
\begin{equation}
\arg[\det(\B')]-\arg[\det(\B)]= \frac{\pi}{2} [({N'_-} -N_-)(\mathrm{mod}\; 2)].
\end{equation}

Comparing with \eref{causticphase}, one verifies that full knowledge of the
signatures  $\sigma (\B)$ and $\sigma(\B')$ only distinguishes the final
$(\pm)$ sign. In this way, the postulation of the Weyl and the chord
representations of metaplectic propagators in terms of determinants without
moduli in \eref{Uweyl} and \eref{Uchord} is justified: The determinants
automatically account for the correct phase changes across caustics, within an
overall sign, which is then specified by \eref{causticphase}.

Finally, one should note that the overall phase increment for crossing a
caustic in the conjugate chord representation, for a continuous evolution that
crosses no Weyl caustic, is obtained in perfect symmetry to the above results.

\section{Product with a translation or a reflection}
\label{sec:producttransrefl}

Operators that correspond to translations or reflections through a point in
phase space are the building blocks of, respectively, the chord and the Weyl
representation \cite{Report,Grossmann,Royer}. Just as the metaplectic
operators, they can also be exactly rendered by semiclassical formulae.
Furthermore, the combination of translations and reflections form the quantum
affine group, which, together with the metaplectic group, integrate the
inhomogeneous metaplectic group \cite{Littlejohn86}, corresponding to the
(classical) inhomogeneous symplectic group. The generating functions for these
transformations add a linear term to the quadratic generating functions of the
(homogeneous) symplectic transformations that we have so far considered
\cite{Report}. 
 
Let us first consider the product of a metaplectic operator $\opU_1$,
represented by $U_1(\x)$ according to \eref{Uweyl} with the unitary
(Heisenberg) translation operator,
\begin{equation}
\hat{T}_{\bxi} = \exp\left(\frac{\rmi}{\hbar}\bxi\wedge\hat{\bi{x}}\right) , \;\;
\textrm{so that}\;\;
T_{\bxi}(\x) = \exp\left(\frac{\rmi}{\hbar}\bxi\wedge\x\right),
\end{equation}
which includes $T_0(\x) = I(\x) = 1$. Assuming that $\opU_1$ is in the
neighbourhood of $\opI$ for which the overall sign is positive, then, according
to \eref{prule}, the product transformation is given by
\begin{equation}
U_{\bxi}(\x) = \pm [\det(\Id + \J\B_1)]^{1/2} \int \frac{\rmd\x_2\rmd\x_1}{(\pi\hbar)^{2N}}
               \exp\!\left(\frac{\rmi}{\hbar} \Phi\right),
\end{equation}
where
\begin{equation}
\Phi = \x_1 \cdot\B_1\x_1 + \x_2 \cdot\J\bxi + 2\x_2 \cdot\J\x_1 
        -2\x_2 \cdot\J\x -  2\x \cdot\J\x_1.
\end{equation}
The integral over $\x_2$ is just a Dirac $\delta$-function,
\begin{eqnarray}
U_{\bxi}(\x) &= \pm [\det(\Id + \J\B_1)]^{1/2} \int \frac{\rmd\x_1}{(\pi\hbar)^{2N}}\\\nonumber
&\hspace{2em}\times(2\pi\hbar)^{2N} \delta(2\x_1 -2\x+\bxi)
\exp\left[\frac{\rmi}{\hbar} (\x_1 \cdot\B_1\x_1-  2\x \cdot\J\x_1)\right],
\end{eqnarray}
so that the result is the inhomogeneous metaplectic transformation,
\begin{equation}
\fl U_{\bxi}(\x) = \pm [\det(\Id + \J\B_1)]^{1/2} 
\exp\left[\frac{\rmi}{\hbar}\big((\x-\bxi/2) \cdot\B_1(\x-\bxi/2)+
       \x \cdot\J\bxi\big)\right],
\label{Uxi}
\end{equation}
that corresponds to the inhomogeneous symplectic transformation generated by
the centre generating function,
\begin{equation} 
S_{\bxi}(\x) = (\x-\bxi/2) \cdot\B_1(\x-\bxi/2)+  \x \cdot\J\bxi,
\end{equation}
through equations \eref{centran}, while holding the translation chord, $\bxi$,
as a fixed parameter. One notes that no new phase appears in the integral over
the $\delta$-function, that is, the overall sign of $ U_1(\x)$ is preserved. In
the limit, $\bxi \rightarrow 0$, i.e. for the product with the identity
operator, it is verified that $[I \cdot U_1](\x) = U_1(\x)$.

The product of a metaplectic operator with the reflection, $\opR_\bi{y}$,
through the phase space point $\bi{y}$ also simplifies, because
\begin{equation}
R_\bi{y}(\x) = (\pi \hbar)^N \delta(\x - \bi{y})
\end{equation}
in the Weyl representation \cite{Report}. Inserting this into \eref{prule} then
leads to
\begin{eqnarray}
U_\bi{y}(\x) &= \pm [\det(\Id + \J\B_1)]^{1/2} \exp\!\left(-\frac{2\rmi}{\hbar} \bi{y} \cdot \J\x\right)
\int \frac{\rmd\x_1}{(\pi\hbar)^{N}}\\\nonumber
&\hspace{5em}\times\exp\!\left[\frac{\rmi}{\hbar} (\x_1 \cdot\B_1\x_1-  2\x_1 \cdot\J(\bi{y}-\x)\right].
\end{eqnarray}
The integral has the same form as the Fourier transform of a complex Gaussian
in the $2N$-dimensional phase space \eref{tanchord} which led to the chord
propagator in \sref{sec:metaplecticops}. So this again leads to a metaplectic
operator with its Weyl symbol:
\begin{equation}
U_\bi{y}(\x) = \pm\rmi^N [\det(\Id + \J{\tilde\B}_1)]^{1/2} 
\exp\!\left(-\frac{\rmi}{\hbar} S_\bi{y}(\x) \right),
\label{Urefl}
\end{equation}
where
\begin{equation}
S_\bi{y}(\x) = (\x-\bi{y}) \cdot{\tilde\B}_1(\x-\bi{y})+  2\bi{y} \cdot\J\x
\label{Sy}
\end{equation}
is the centre generating function for the corresponding inhomogeneous
symplectic transformation specified by \eref{centran}. In the case of a
reflection through the origin (i.e. a parity operator), $\bi{y}=0$ and
\eref{Urefl} will be just the homogeneous metaplectic operator specified by
$-{\tilde\B}_1$, i.e. the product propagator adopts the Cayley matrix of the
original chord propagator.

The factor $\rmi^N$ may seem strange, since it does not fit in with the phases
in \eref{Uweyl}.  If one chooses $U_1(\x)$ for a harmonic osillator, the
product may be matched to just a further rotation in phase space by $\pi/2$, so
that presumably \eref{Urefl} should satisfy \eref{Uho}.  Thus, one should
recall the conclusion in \sref{sec:metaplecticops} that the appropriate
reflection included in the metaplectic group is not $\hat{R}_{\x}$, but the
operator $\hat{R}'_{\x}= \rmi^{-N}\hat{R}_{\x}$.  For the product of this
metaplectic reflection with a general metaplectic operator, the factor of
$\rmi^N$ is indeed cancelled in the corresponding expression to \eref{Urefl}.
In applications where the product arises because evolution acts on an operator
specified by its Weyl symbol, this extra factor must be included, because the
reflection operator implied will then indeed be $\hat{R}_\x$, rather than
$\hat{R}'_{\x}$.

In practice, the phase increment due to the product by $\hat{R}_\x$ is just 
\begin{equation}
\Theta=\frac{\pi}{4} \sigma(\B_1)=\frac{\pi}{2}(N-N_{1-}) ,
\end{equation}
where $N_{1-}$ is the number of negative eigenvalues of $\B_1$.  This
determines the $(\pm)$ sign and it is exactly the same phase increment as
previously obtained in passing from the Weyl propagator to the chord propagator
in \sref{sec:metaplecticops}.

For the sake of brevity, we note that in the chord representation products of
translations or reflections with metaplectic operators are just Fourier
transforms of \eref{Uxi} and \eref{Urefl}.  Basically, the corresponding Cayley
matrices for factors and their products are given by \eref{BB}.  Here again,
one must carefully choose between $\hat{R}_\x$ and $\hat{R}'_{\x}$, depending
on the context: their chord symbols are Dirac $\delta$-functions with different
phase factors.

\section{Products of metaplectic transformations}
\label{sec:productsmetaplectic}

We now consider general products of pairs of arbitrary metaplectic
transformations, $\opU = \opU_2\opU_1$, corresponding to symplectic
transformations, such that $\M = \M_2\M_1$, parametrized by Cayley matrices,
$\B_1$ and $\B_2$, according to \eref{Cayley1}, so that both factors take on
the form \eref{Uweyl}.  For simplicity, we shall assume that both $\opU_1$ and
$\opU_2$ have overall positive signs.  Inserting these ingredients into the
product rule \eref{prule} then leads to
\begin{equation}
\fl U(\x) = [\det(\Id + \J\B_1)]^{1/2}  [\det(\Id + \J\B_2)]^{1/2}
\int \frac{\rmd\x_2\rmd\x_1}{(\pi\hbar)^{2N}} \exp\!\left(\frac{\rmi}{\hbar} \Phi(\x, \x_1, \x_2)\right),
\end{equation}
where
\begin{equation}
\fl \Phi(\x, \x_1, \x_2) = \x_1 \cdot\B_1\x_1 + \x_2 \cdot\B_2\x_2 
+ 2\x_2 \cdot\J\x_1 -  2\x_2 \cdot\J\x -  2\x \cdot\J\x_1
\end{equation}
is a homogeneous quadratic form in its $6N$ variables, or an inhomogeneous
quadratic form in the $4N$ variables $(\x_1, \x_2)$ if we fix the centre $\x$
where the full propagator is evaluated.  It is shown in \cite{Report} that,
upon obtaining $\x_1(\x)$ and $\x_2(\x)$ from the conditions,
$\partial\Phi/\partial\x_1=0$ and $\partial\Phi/\partial\x_2=0$, the resulting quadratic form
in $\x$ is just the centre generating function for the product transformation,
that is, $\Phi(\x, \x_1(\x), \x_2(\x))=S(\x)=\x \cdot\B\x$, corresponding to
$\M=\M_2\M_1$.  Adopting this stationary point of $(\x_1, \x_2)$ as the origin
then leads to
\begin{eqnarray}
\label{Uprod}
U(\x) &= [\det(\Id + \J\B_1)]^{1/2}  [\det(\Id + \J\B_2)]^{1/2} 
\exp\!\left(\frac{\rmi}{\hbar}\x \cdot\B\x\right)\\\nonumber
&\hspace{10em}\times\int \frac{\rmd\x_2\rmd\x_1}{(\pi\hbar)^{2N}}
\exp\!\left(\frac{\rmi}{\hbar} \Phi(0, \x_1, \x_2)\right),
\end{eqnarray}
so that our task is to determine the amplitude and phase for the $4N$-dimensional 
complex Gaussian integral in \eref{Uprod}.

One must evaluate the determinant of the $4N$-dimensional matrix
$[\doubleB-\doubleJ]$, where one defines the symmetric block matrices:  
\begin{equation}
\label{doublemat}
\doubleB \equiv  
\left(
\begin{array}{cc}
     {\B_1} & 0     \\
      0     & {\B_2} 
\end{array}
\right) 
\;\;{\rm and}\;\;
\doubleJ \equiv
\left(
\begin{array}{cc}
      0      & {\J} \\
      {-\J}  &  0 
\end{array}
\right) \; .
\end{equation}
The phase increment is then 
\begin{equation}
\Theta = \frac{\pi}{4}\sigma(\doubleB-\doubleJ) = \pi\left(N -\frac{\doubleN_-}{2}\right),
\label{phinc}
\end{equation}
where $\doubleN_-$ is the number of negative eigenvalues.  In the limit
$\doubleB \rightarrow 0$, one is left with $-\doubleJ$, which has $2N$
eigenvalues $+1$ and $2N$ eigenvalues $-1$, so that $\det(\doubleB-\doubleJ) =
1$ and $\sigma(\doubleB-\doubleJ)=0$.  This confirms a basic assumption for
previous results in this paper: There is no overall change of sign for the
product transformation, as long as the matrix $\doubleB$ can be treated as a
small perturbation of $\doubleJ$.  Indeed, for a continuous  one-parameter
family of matrices $\doubleB$, it is only at $\det(\doubleB-\doubleJ) = 0$, as
an eigenvalue vanishes, that there may occur an overall change of phase, unless
either of the matrices $\B_j$ becomes singular.

Just as in the Fourier transform between the Weyl and the chord propagators in
\sref{sec:metaplecticops}, the phase of the determinant of the relevant
quadratic form already captures part of the information supplied by its
signature, i.e.  $\arg[\det(\doubleB-\doubleJ)]=\pi[\doubleN_-(\mathrm{mod}\;
2)]$.  The block form of the $4N$-dimensional determinant allows for the
simplification:
\begin{equation}
\det(\doubleB-\doubleJ) = (-1)^{2N} \det
\left(
\begin{array}{cc}
      {-\J}  & {\B_1}  \\
      {\B_2} & {\J} 
\end{array}
\right) 
= \bDelta,
\label{detDelta}
\end{equation}
where we define,
\begin{equation}
\bDelta \equiv \det(\Id+ \J\B_2 \J\B_1) = \det(\Id+ \J\B_1 \J\B_2) .
\label{Delta}
\end{equation}
Further understanding of the $2N$-dimensional matrix with this determinant is
gained by drawing back to the interpretation of the amplitude of the Weyl
propagators in terms of the matrix \eref{initialmid} between the midpoint and
the endpoint of the classical trajectory. For the product transformation with
symplectic matrices $\M=\M_2\M_1$, we then have
\begin{equation}
\frac{1}{2}(\Id+\M) = (\Id+ \J\B_2)^{-1}(\Id+ \J\B_2 \J\B_1)(\Id+ \J\B_1)^{-1}.
\label{prod1}
\end{equation}
The alternative form of \eref{Delta} arises similarly from the inverse product
transformation to \eref{prod1}, i.e. $\M^{-1}={\M_1}^{-1}{\M_2}^{-1}$ is related
to 
\begin{equation}
\frac{1}{2}(\Id+{\M}^{-1}) = (\Id- \J\B_1)^{-1}(\Id+ \J\B_1 \J\B_2)(\Id - \J\B_2)^{-1}.
\label{prod3}
\end{equation}
Hence, the amplitude of the Weyl propagator for the product operator follows from
\begin{equation}
2^{2N} \det(\Id+\M)^{-1} = \det(\Id+ \J\B_2)\det(\Id+ \J\B_1)\bDelta^{-1}.
\label{prod2}
\end{equation}

The conclusion is that the product of a pair of metaplectic operators in the
Weyl representation \eref{Uweyl} results in a Weyl propagator of exactly the
same form, but expressed in terms of the Cayley matrix for the combined
symplectic transformation. The indeterminacy of the overall sign in the square
root of the determinantal amplitude can only be lifted by evaluating the phase
$\Theta$ in \eref{phinc}.  In the case of a continuous one-parameter family of
product transformations, i.e. a one-parameter family of double matrices
$\doubleB$ (or, in the simplest case, a one-parameter family of matrices
$\B_j$, with the other matrix held fixed) a possible change of the overall
phase of the product only arises if there is a change of sign of
$\det(\doubleB-\doubleJ)=\bDelta$.  But, according to \eref{prod1},
this event entails a change of sign for the final $\det(\Id+\M)$, itself.  So
there is no change of phase unless the combined symplectic transformation
crosses a caustic.

It is often the case that $\opU_2$ can be identified with a member of a family
$\opU_t$ for $t=t_2$, while $\opU_{t=0}=\opI$; for instance, the symplectic
Hamiltonian flow: $\M = \exp[t \J\mathbf{H}]$, where $\mathbf{H}$ is the
Hamiltonian matrix.  Here, $\opU_1$ is an arbitrary passive metaplectic
operator.  Then, if $\det(\M_t\M_1-\Id)\neq 0$ from $t=0$ until $t=t_2$, the
phase increment for a product can be obtained from the difference of signatures
of $2N$-dimensional matrices \eref{causticphase}.  One must be wary of the
change of notation, since the transition of the Cayley matrices in
\sref{sec:generalizedmaslov}, which was $\B \mapsto \B'$, becomes here $\B_1
\mapsto \B$.

The advantage of this alternative approach relying on continuity is that it
only deals with $2N$-dimensional matrices. In the case of the chord
representation, even the products of metaplectic operators can be considered
within a $2N$-dimensional integral, because of its simpler product rule.  The
deduction of the double sheet structure of metaplectic operators has previously
been achieved by Littlejohn \cite{Littlejohn86} within the position
representation and for the chord representation (with different notation) by de
Gosson \cite{deGosson06}.  Our object here is not merely to show how the group
of metaplectic transformations is represented by Weyl symbols, but above all to
determine the phase increment in each specific case. 

It may happen that $\bDelta$ and hence $\det(\doubleB-\doubleJ)$ has a
double root for a family of operator products. An important instance of this
occurs if the commutator $[\J\B_2, \J\B_1]=0$. This includes the case where
$\M=(\M_1)^2$, or where $\M_1$ and $\M_2$ are generated by the same quadratic
Hamiltonian.  Then $\J\B_1$ and $\J\B_2$ can be simultaneously diagonalized,
such that $\lambda_1^{(k)}$ and $\lambda_2^{(k)}$ are their eigenvalues for the
common $k$-th eigenvector of both these matrices.  Hence, the caustic condition
for the product transformation, $\bDelta=0$, reduces to $\lambda_1^{(k)}
\lambda_2^{(k)} = -1$. This condition allows for elliptic transformations, but
not direct hyperbolic transformations, for which $|\lambda_j|<1$. On the other
hand, $-\lambda_j^{(k)}$ will also be a pair of simultaneous eigenvalues, so that
$\bDelta$ has a double root.  The change of phase at the caustic is
then an integral multiple of $\pi$, which is consistent with the need for a
product of elliptic transformations to remain elliptic, i.e. there can be no
transition to hyperbolic with reflection. 
 
An example of a single null eigenvalue occurs at the caustic of a family of
products of elliptic with hyperbolic transformations:
\begin{eqnarray}
\label{elhyp1}
\B_1 = 
\omega \left(
\begin{array}{cc}
     1 & 0 \\
     0 & 1 
\end{array}
\right) \;, \;
\B_2 = 
\gamma \left(
\begin{array}{cc}
     1 & 0 \\
     0 & -1 
\end{array}
\right)\\\nonumber
\Id + \J\B_1\; \J\B_2 = 
\left(
\begin{array}{cc}
     1+\gamma\omega & 0 \\
     0              & 1-\gamma\omega 
\end{array}
\right) .
\end{eqnarray}
Thus, defining the one-parameter family of transformations either by the
parameter $\gamma$ or by the parameter $\omega$, there will be a single null
eigenvalue, i.e. single zero of $\bDelta$, at $\gamma\omega =1$ and
another one at $\gamma\omega =-1$. The symplectic matrices corresponding to
\eref{elhyp1} are
\begin{equation}
\label{elhyp2}
\fl \M_1 = 
\frac{1}{1+\omega^2} \left(
\begin{array}{cc}
     1-\omega^2 & 2\omega \\
     -2\omega   & 1-\omega^2 
\end{array}
\right) \;, \;
\M_2 = 
\frac{1}{1-\gamma^2} \left(
\begin{array}{cc}
     1+\gamma^2 & -2\gamma \\
     -2\gamma   & 1+\gamma^2 
\end{array}
\right)\;.
\end{equation}
For the caustic condition $\gamma\omega=1$, one obtains 
\begin{equation}
\label{elhyp3}
\Id+\M = 
\frac{4}{\omega^2-\gamma^2} \left(
\begin{array}{cc}
     1         & \omega \\
     -\gamma   & -1 
\end{array}
\right) \;, 
\end{equation}
so that $\det(\Id+\M) = \tr(\M)+2=0$, but $\M$ is not a reflection.  One
can check that $\det(\Id+\M)$ does change its sign as the product parameter
$\gamma\omega$ passes the value 1, which leads to a transition of the type of
symplectic transformation for the product.

In all cases, full account must be taken of continuity for the proper family of
tangent propagators.  Indeed, one or both of the factor families may cross a
caustic, without any effect on the overall phase of the product. An important
instance is the {\it fidelity}, or {\it Loschmidt echo operator} treated in
\cite{OAValZam}, such that the product of finite evolutions remains close to
the identity operator.  Continuity of the product operator constrains the
continuity of its Weyl representation, in spite of any possible discontinuities
in the phase of the factor propagators.  This may considerably simplify
practical calculations.  The way that the phase jumps cancel for the particular
example of a product of harmonic oscillator evolutions is analyzed in the
Appendix.

It should be remarked that the SC propagator for Wigner functions may be
considered as a special case of a compound propagator resulting from the
product of evolution operators.  This has the peculiarity that the dominant
classical trajectory is precisely on a caustic, so that a higher uniform
approximation has been developed, for which the phases were previously analyzed
\cite{Dittrich,DitGoPa}. In contrast, the mixed propagator presented in
\cite{AlmBro06} avoids the caustic, so that its tangent propagator for short
times is given by \eref{Uweyl} with a positive sign.  It is shown in
\cite{OAValZam} that there are important applications where the caustic
disappears within an integration, so that one needs only worry about the phase
increment, which is the focus of the present study.

\section{Conclusions}
\label{sec:conclusions}

Notwithstanding the theoretical interest in the way that the Weyl symbol
renders the double sheeted metaplectic group, practical use of this phase space
representation requires clear rules for phase increments. Indeed, one should
not need to be reminded about the abstract topology of this group so as to
evolve quantum operators in the SC approximation. The difficulty lies in the
crossing of caustics, either as time changes (while the argument of the Weyl
propagator is fixed), or for different arguments of the same Weyl propagator.
In both cases, the phase increment coincides with that of the appropriate
family of metaplectic operators, each of which corresponds to a definite
classical trajectory.  Fortunately, one can indeed summarize the conclusions in
a nutshell:  

\begin{enumerate}
\item Continuity of the family of metaplectic operators, corresponding to the
family of tangent maps neighbouring each classical trajectory, guarantees that
there will be no SC phase jump, unless a caustic is crossed. This even holds
for compound trajectories that are pieced together to form {\it super
propagators} for Wigner functions, or for SC evaluations of the quantum
fidelity. No matter how many caustics may have been traversed by the factor
operators, one only needs to worry about phase jumps if the product propagator
has crossed a caustic. In a causticless neighbourhood of the origin, the sign
for the tangent propagators in \eref{Uweyl} is positive.

\item For each caustic traversal of the Weyl propagator (compound or not), one
needs a sample of the classical Cayley matrix for a pair of arbitrary instants
before and after this SC singularity is met.  This requires, either the
quadratic approximation of the centre action (i.e. the centre generating
function \eref{centaction}) or else the relation to the (linearized) symplectic
transformation \eref{Cayley-1}.  Then the phase jump is obtained by the
difference in the signature between these Cayley matrices \eref{causticphase}.
By expressing the amplitude of the Weyl propagators in terms of a determinant
without a modulus in \eref{Uweyl}, any factor of $\rmi$ is already accounted for
beyond the caustic. The remaining ambiguity concerning the final phase of $\pi$
is then resolved by adding the above phase to the phase preceding the caustic
traversal. The required information is naturally available in any SC computation.
\end{enumerate}

Examples of these rules for products of propagators corresponding to (exactly
metaplectic) harmonic oscillators with different frequencies are discussed in the
appendix. 

The symmetry between Weyl propagators and their Fourier transform leads to
similar phase recipes for chord propagators.  They are an equally valuable
source of super operators for which one must determine the precise phase.  The
phase jump upon crossing a caustic is obtained by simply replacing $\B$ and
$\B'$ in \eref{causticphase} by $-\tilde{\B}$ and $-\tilde{\B}'$.

\appendix

\section{Product of two oscillations}

Let us consider the product of evolution operators for a pair of harmonic
oscillators $(N=1)$, described by the Weyl propagators \eref{Uho}. It should be
recalled that these may be considered as tangent propagators for general
nonquadratic Hamiltonians, such that the corresponding classical tangent maps
are both elliptic. The simplifying assumption here is that both maps can be
diagonalized simultaneously. Conveniently one keeps a common time, $t$, as a
single parameter for the product, while the frequency $\omega_2$ may differ
from $\omega_1$, even with respect to its sign, denoting a different sense of
oscillation. The Weyl propagator for the product is then 
\begin{eqnarray}
\label{Uhopro}
U_t(\x) &= \frac{\exp \left[-\frac{\rmi}{\hbar} \tan \left(\frac{\omega_1 + \omega_2}{2} t\right) \x^2 \right]}
{\cos \left(\frac{\omega_1}{2} t\right) \cos \left(\frac{\omega_2}{2} t\right)}
\int \frac{\rmd\x_2\rmd\x_1}{(\pi \hbar)^{2}}\\\nonumber
&\hspace{5em}\times\exp\left[\frac{\rmi}{\hbar}(\Omega_1 {\x_1}^2 +\Omega_2 {\x_2}^2 + 2\x_2 \cdot \J\x_1)  \right].
\end{eqnarray}
where
\begin{equation}
\Omega_j= -\tan\left(\frac{\omega_j t}{2}\right).
\end{equation}
The exponent in the above integral is just the quadratic form determined by the
symmetric matrix, $\doubleB-\doubleJ$, for which 
\begin{equation}
\det(\doubleB-\doubleJ) = \bDelta = (\Omega_1\Omega_2-1)^2
= \frac{\cos \left(\frac{\omega_1 + \omega_2}{2} t\right)}
{\cos \left(\frac{\omega_1}{2} t\right) \cos \left(\frac{\omega_2}{2} t\right)},
\label{A3}
\end{equation}
while the characteristic equation,
\begin{equation}
\det(\doubleB-\doubleJ-\lambda\doubleId) = [(\Omega_1-\lambda)(\Omega_2-\lambda)-1]^2 = 0,
\end{equation}
has double roots:
\begin{equation}
\lambda_\pm = -\frac{\Omega_1+\Omega_2}{2}
\pm\left[1 + \left(\frac{\Omega_1-\Omega_2}{2}\right)^2\right]^{\frac{1}{2}}.
\end{equation}
Hence, there are two alternatives for the overall change of phase increment \eref{phinc},
  
\begin{enumerate}
\item $\Theta=0$ if $\Omega_1\Omega_2<1$.
\item $\Theta=\pm \pi$ if $\Omega_1\Omega_2>1$.
\end{enumerate}

The simplest case is that of a product of forward oscillations ($\omega_1$ and
$\omega_2>0$) such that neither has individually reached the caustic at
$\omega_j t=\pi$, but $(\omega_1+\omega_2)t>\pi$. This is just case (ii), so
that, if both $\omega_j t\approx\pi$, then $U_t(\x)\approx -1$.

For the Loschmidt echo, the returning motion is specified by $\omega_2 \approx
-\omega_1$, so that $\Omega_2\approx -\Omega_1$. For short times, both
$\cos(\omega_j t/2)>0$, so that $\Omega_1\Omega_2<0$, which is case (i).  There
is no new phase and we obtain $U_t(\x)\approx 1$.  If one now allows $\omega_1$
to be slightly larger than $\pi$, though still $|\omega_2 t|<\pi$, then
$\Omega_1>0$ and $\Omega_1\Omega_2>1$, which is case (ii) and hence
$\Theta=\pi$.  But this exactly cancels the phase coming from $\cos(\omega_1
t/2)<0$, so that the overall sign of the propagator remains positive, just as
for small times, that is, $U_t(\x)\approx 1$. Finally, if one allows both
oscillators for the Loschmidt echo to pass their caustic, we are back in case
(i) since $\Omega_1\Omega_2<0$. But now both amplitudes, $\cos(\omega_j
t/2)<0$, as long as $|(\omega_1 + \omega_2)t|<\pi$, so the overall sign is
still positive and, once again, $U_t(\x)\approx 1$.

The special simplicity of the harmonic oscillator allows for an appealing
though untypical synthesis of these results, due to the identification of the
previous conditions with the sign of $\bDelta^{1/2}$ in \eref{A3}: 

\begin{enumerate}
\item $\cos \left(\frac{\omega_1}{2} t\right) \cos \left(\frac{\omega_2}{2}
t\right)>0$; then the sign coming from the integration is a minus if $\cos
\left(\frac{\omega_1 + \omega_2}{2} t\right)<0$, but this converts $\left|\cos
\left(\frac{\omega_1 + \omega_2}{2} t\right)\right|$ into $\cos
\left(\frac{\omega_1 + \omega_2}{2} t\right)$.

\item $\cos \left(\frac{\omega_1}{2} t\right) \cos \left(\frac{\omega_2}{2}
t\right)<0$; then the integration sign is negative if $\cos
\left(\frac{\omega_1 + \omega_2}{2} t\right)>0$, which converts $\cos
\left(\frac{\omega_1}{2} t\right) \cos \left(\frac{\omega_2}{2} t\right)<0$
into $\left|\cos \left(\frac{\omega_1}{2} t\right) \cos
\left(\frac{\omega_2}{2} t\right)\right|$.
\end{enumerate}

\begin{figure}
\centering
\includegraphics[width=0.45\textwidth]{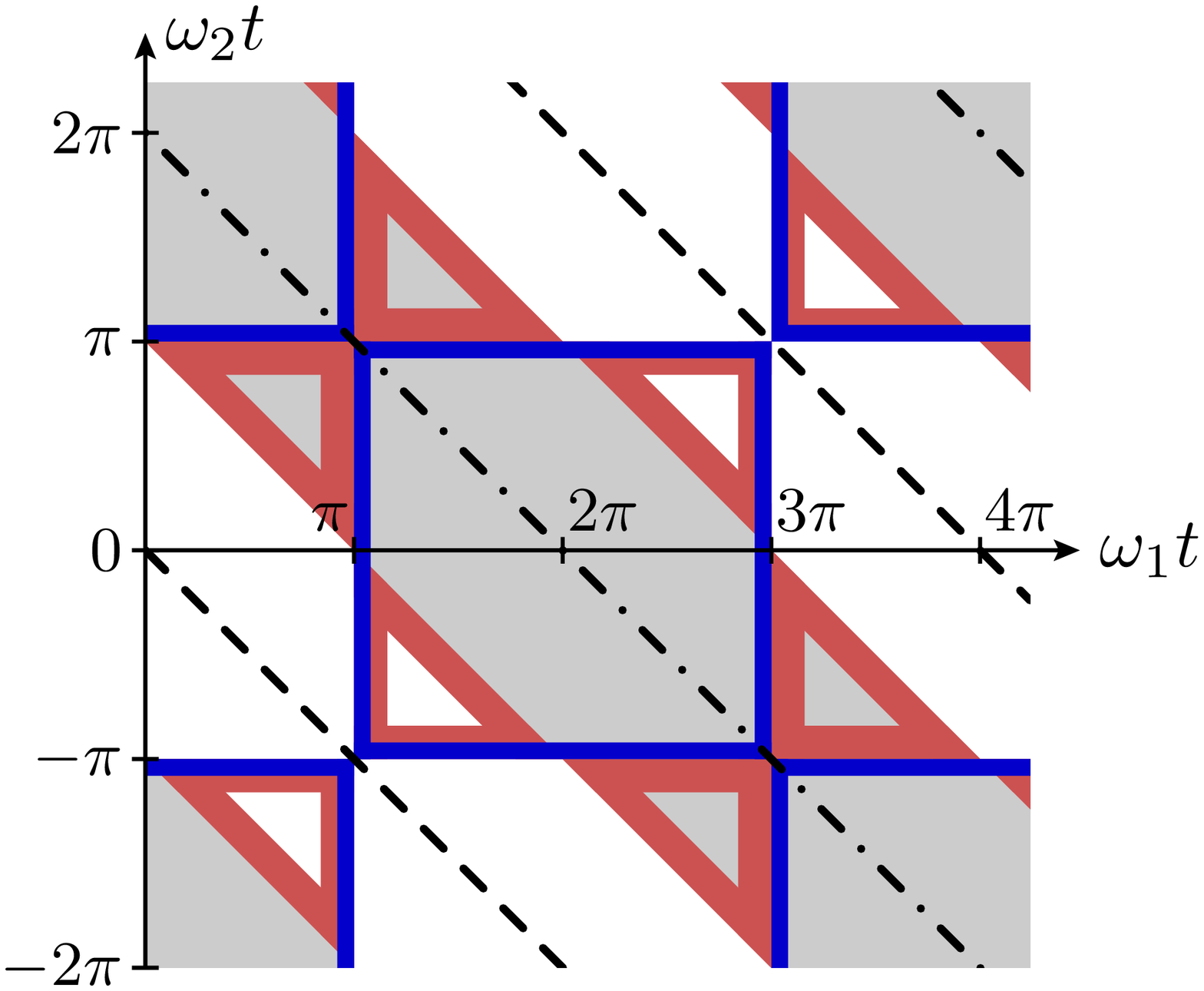}
\caption{Overall sign for a product of harmonic oscillations: A negative sign
arises within the blue (black) squares and within the red (dark grey)
triangles, due to the sign of the product of the initial cosines or the
crossing of a caustic, respectively. Together these generate a negative sign
for the product propagator  in the interior of the light grey stripes. The
dashed lines correspond to the unit operator $\opI$, while all points on the
dash-dotted lines correspond to $-\opI$.}
\label{Fig2}
\end{figure}

In all cases the amplitude will be $\cos \left(\frac{\omega_1 + \omega_2}{2}
t\right)$ without any sign ambiguity, as portrayed in \fref{Fig2}. One
should be warned that this is a very special case: In general, one cannot
evaluate the final phase without keeping detailed track of the intial phases
and the phase increment!

For a Loschmidt echo, the product of oscillations as a function of time will
define a vector in \fref{Fig2} that grows from the origin, nearly parallel to
the dashed line. Thus, it will be a long time before it finally leaves the
white strip and changes its sign. In contrast, the pair of oscillations in the
same sense, that were first considered, runs in the direction of the other
diagonal, so that the sign changes repeatedly as $t$ increases.

\ack
We thank Raul Vallejos, Eduardo Zambrano and  Ji\v{r}\'{\i} Van\'{\i}\v{c}ek
for stimulating discussions. AMOA thanks the hospitality of the University of
Augsburg and, reciprocally, GLI is greatful to CBPF for its hospitality.
Partial financial support from the National Institute for Science and
Technology--Quantum Information, FAPERJ and CNPq (Brazilian agencies) is
gratefully acknowledged.

\end{document}